\documentclass[10pt]{article}
\usepackage{amssymb}
\usepackage{amsmath}
\usepackage{amsthm}
\usepackage{latexsym}
\usepackage[dvips]{epsfig}
\usepackage{mathrsfs}
\usepackage{eufrak}

\theoremstyle{plain}
\newtheorem{proposition}{Proposition}

\newtheorem*{conjecture}{Conjecture}
\newtheorem*{subconjecture}{Subconjecture}

\font\SYM=msbm10
\newcommand{\Real}{\mbox{\SYM R}}
\newcommand{\Complex}{\mbox{\SYM C}}
\newcommand{\Natural}{\mbox{\SYM N}}

\font\tenscr=rsfs10 scaled1100
\font\sevenscr=rsfs7 
\font\fivescr=rsfs5 
\skewchar\tenscr='177
\skewchar\sevenscr='177
\skewchar\fivescr='177
\newfam\scrfam
\textfont\scrfam=\tenscr
\scriptfont\scrfam=\sevenscr
\scriptscriptfont\scrfam=\fivescr

\def\scri{{\fam\scrfam I}}

\def\O{\mathcal{O}}

\newcommand{\TT}[3]{T_{#1 \phantom{#2} #3}^{\phantom{#1} #2}}
\newcommand{\updn}[3]{#1^{#2}_{\phantom{#2}#3}}
\newcommand{\dnup}[3]{#1_{#2}^{\phantom{#2}#3}}

\begin{document}


\title{\textbf{The Maxwell field on the Schwarzschild spacetime: behaviour near spatial infinity}}

\author{{\Large Juan Antonio Valiente Kroon} \thanks{E-mail address:
 {\tt j.a.valiente-kroon@qmul.ac.uk}} \\
School of Mathematical Sciences, Queen Mary, University of London,\\
Mile End Road, London E1 4NS, United Kingdom.}

\maketitle

\begin{abstract}
  The behaviour of the Maxwell field near one of the spatial
  infinities of the Schwarzschild solution is analysed by means of the
  transport equations implied by the Maxwell equations on the cylinder
  at spatial infinity. Initial data for the Maxwell equations will be
  assumed to be expandable in terms of powers of a coordinate $\rho$
  measuring the geodesic distance to spatial infinity (in the
  conformal picture) and such that the highest possible spherical
  harmonics at order $p$ are $2^p$-polar ones. It is shown that if the
  $2^p$-polar harmonics at order $p$ in the initial data satisfy a certain 
  regularity condition then the solutions to the transport equations
  at orders $p$ and $p+1$ are completely regular at the critical sets
  where null infinity touches spatial infinity. On the other hand, the
  solutions to the transport equations of order $p+2$ contain, in
  general, logarithmic singularities at the critical sets. It is
  expected that the ideas and techniques developed in the analysis of
  this problem can be employed to discuss the more challenging case of
  the transport equations implied by the conformal Einstein equations,
  and thus provide a way of constructing a proof of a certain rigidity
  conjecture concerning the developments of conformally flat initial
  data sets.

\end{abstract}



\section{Introduction}
In \cite{Val04a} the following rigidity conjecture concerning the
development of conformally flat initial data for the Einstein vacuum
equations sets was put forward:

\begin{conjecture} \label{conjecture} 
  If an initial data set for the
  Einstein vacuum equations which is time symmetric and conformally
  flat in a neighbourhood $\mathcal{B}_a(i)$ of infinity yields a
  development with a smooth null infinity, then the initial data set
  is Schwarzschildean in $\mathcal{B}_a(i)$.
\end{conjecture}

There is further evidence for generalisations of this conjecture to
more general classes of initial data ---see e.g.
\cite{Val04d,Val04e,Val05a}. A presumptive proof of such a
conjecture would entail two aspects:

\begin{itemize}
\item[(I)] A deep understanding of the mechanisms of the asymptotic
  expansions that are obtained using Friedrich's construction of the
  cylinder at spatial infinity ---see e.g. \cite{Fri98a, Fri04}.

\item[(II)] The construction of estimates which allow to get around
  the manifest degeneracy of the Einstein conformal equations at the
  critical sets were null infinity ``touches'' spatial infinity ---see
  e.g. \cite{Fri03b}.
\end{itemize}
 
The present article explores a set of ideas on how to deal with the 
aspect (I) mentioned above.

\bigskip 
Friedrich's seminal work on the \emph{$i^0$ problem} has
rendered a representation of spatial infinity which exploits to the
maximum the conformal properties of the spacetime by choosing a
convenient gauge.  In particular, spatial infinity is represented in
this gauge by an extended set, the cylinder at spatial infinity,
$\mathcal{I}$ ---see \cite{Fri98a}. This framework uses the
\emph{extended conformal field equations} ---as given in, say,
\cite{Fri95,Fri98a}--- to pose a \emph{regular initial value problem
  at spatial infinity}.

The cylinder at spatial infinity can be regarded as a limit set of
incoming and outgoing light cones. It has the remarkable
property of being a \emph{total characteristic} of the conformal
Einstein propagation equations. This property implies that the
propagation equations reduce in their entirety to an interior system
of transport equations upon evaluation at $\mathcal{I}$. This allows
to \emph{transport} information from some initial data prescribed on
some initial hypersurface $\mathcal{S}$ up to the critical sets
$\mathcal{I^\pm}$ where spatial infinity touches null infinity. Let
$\phi=(\phi_0,\phi_1,\phi_2,\phi_3,\phi_4)$ denote the components of
the rescaled Weyl tensor in a particular gauge and with respect to a
particular frame ---to be described in the subsequent sections. Then
the transport equations at $\mathcal{I}$ allow to calculate
$\partial_\rho^{p}\phi|_\mathcal{I}$ ---where $\rho$ is some radial
coordinate measuring the distance to $\mathcal{I}$--- on the whole of
$\mathcal{I}$ ---and in particular at $\mathcal{I}^\pm$--- if the
value of $\partial_\rho^{p}\phi$ on
$\mathcal{I}^0=\mathcal{I}\cap\mathcal{S}$ is given and the lower
order derivatives $\partial^k_\rho\phi|_\mathcal{I}$, $k=0,\ldots,p-1$
are known. This unfolding of the evolution process can be thought of
as the construction of the Taylor like expansions of the form
\begin{equation} \label{expansion}
\phi\sim \sum_p \frac{1}{p!}\left(\partial_\rho^p \phi|_\mathcal{I}\right) \rho^p.
\end{equation}
A general observation is that the derivatives $\partial^p_\rho
\phi|_\mathcal{I}$ contain logarithmic divergences at
$\mathcal{I}^\pm$. By logarithmic divergences at, say $\tau=\pm1$,  it will be understood the appearance expressions of the form
\begin{equation}
(1-\tau)^{k_-}\ln(1-\tau), \quad (1+\tau)^{k_+}\ln(1+\tau).
\end{equation} 
Note how the exponents $k_-$ and $k_+$ regulate the smoothness of the
term. The logarithmic divergences appearing in the solutions of the
transport equations implied by the conformal Einstein equations can be
grouped in two types:

\begin{itemize}
\item[(i)]those appearing from general linear aspects of the equations;

\item[(ii)]and those arising from the particular nonlinear aspects of the
equations.
\end{itemize}

The logarithmic divergences of (i) are well understood and their
preclusion imposes mild conditions on the initial data ---see
\cite{Fri98a,Val03a}. Not surprisingly, it is the understanding of the
logarithmic divergences of type (ii) which poses the real challenge.
The discovery of a particular class of this logarithms led to
conjecture \ref{conjecture}. At the moment, there are no general
results regarding the behaviour of this kind of logarithms. The only
knowledge available is what has been inferred from the calculation of
the expansions (\ref{expansion}) up to a finite order ---say
$p=7,8$--- using computer algebra methods.

With regards to the coefficients
$(\partial^p_\rho\phi|_{\mathcal{I}})$ in the expansions
(\ref{expansion}), one would like to prove a result on the lines of:

\begin{subconjecture}
  Given a time symmetric initial data which is conformally flat on
  $\mathcal{B}_a(i)$, and if $\partial_\rho^p\phi|_\mathcal{I}$ are smooth
  at $\mathcal{I}^\pm$ for all $p$, then
  $\partial_\rho\phi|_{\mathcal{I}^0}$ is Schwazschildean.
\end{subconjecture}

In other words, if one starts with a general time symmetric,
conformally flat initial data set and one works out the 
behaviour of the derivatives $\partial^p_\rho\phi|_\mathcal{I}$ at
$\mathcal{I}^\pm$, then one will observe that generically there are
logarithmic terms. If one removes the parts of the initial data
giving rise to these divergences at every order $p$, then one will
observe that the jet $\{\partial_\rho\phi|_{\mathcal{I}^0},
\;p=0,1,\ldots \}$ is equal to the one obtained from looking at the
$t=constant$ slice of the Schwarzschild solution in isotropic
coordinates. In order to make these ideas work, one needs the
understanding of the asymptotic expansions which was alluded to in point
(I).

The complexity and vast number of equations which has to be controlled
when dealing with the conformal Einstein equations makes developing
the required understanding a daunting endeavour. That is the reason
why it is desirable to have simpler ``toy models'' on which diverse
techniques could be tested and from which underlying structures could
be apprehended. Devising such a toy model is a delicate issue. Of
course, one wishes for a tractable model, but at the same time one
would like to avoid simplifying things too much and render the
analysis trivial.

The obvious ``toy model'' in the form of a spin-2 zero-rest-mass field
propagating close to one of the spatial infinities of the
Schwarzschild spacetime is not an adequate one in view of the
existence of an obstruction in the form of the Buchdahl constraint
which precludes the consideration of generic fields in the analysis
---see e.g. \cite{Buc58}. This problem is overdetermined.

Due to the above considerations, attention is turned to another
possible toy model: the propagation of the Maxwell field (spin-1
zero-rest-mass field) in the vicinity of one of the spatial infinities
of the Schwarzschild spacetime. A direct application of the ideas of,
say, \cite{Val03a} shows the existence of logarithmic divergences at
the sets $\mathcal{I}^\pm$ of the underlying Schwarzschild spacetime
which are due to generic linear features of the Maxwell equations
which are shared with the conformal Einstein equations ---the type (i)
divergences. An understanding of the mechanism behind these
divergences, as it will be seen, poses no challenge. The real
difficulties arise from the existence of another type of logarithmic
divergences which are generated by the interaction of the Maxwell
field with the background spacetime. A preliminary analysis of the
analogue of the expansions (\ref{expansion}) for this toy model using
the computer algebra methods of \cite{Val04a} reveals that the inner
workings of the logarithmic divergences are similar to those of the
nonlinear propagation of the gravitational field.

The main goal of the present work is to develop a procedure of
tracking down which parts of the initial data are responsible for
specific logarithmic divergences at the critical sets. To this end it
shall be assumed that one has a class of initial data for the Maxwell
equations expandable in powers of a geodesic distance $\rho$ to
infinity. Further, it will be assumed that the coefficients of this
expansion at say, order $\rho^p$ admit at most $2^p$-polar harmonics ---this
happens if, for example, the initial data is analytic in a neighbourhood of
infinity. Consistent with the discussion of the previous paragraphs,
it will also be assumed that the initial data satisfies the regularity
conditions precluding the appearance of logarithms of the so-called
type (i). Given this class of initial data, the question is the
following: at which order ---if at all--- do the $2^p$-polar terms in
the initial data produce divergences at the critical sets? Transport
equations for $2^p$-polar quantities appear for the first time at
order $p$. The present analysis shows that the solutions at orders $p$
and $p+1$ do not contain logarithmic divergences ---for order $p$ this is
direct, but for order $p+1$ there is an intriguing ``conspiracy'' of terms
taking place. Moreover, for order $p+2$ the solutions contain, in
general, the logarithmic terms. A similar pattern ---although with a
further layer of complication--- arises for the conformal Einstein field
equations. Hence the interest of the present analysis.

It should be mentioned that there has recently been  some related work
on the peeling properties of solutions of the wave equation on a
Schwarzschild background ---see \cite{MasNic04,MasNic07}. These
analyses can, most certainly, be extended to the case of the Maxwell
equations. In any case, the scope, aims, techniques and motivations for 
these results are different to the ones in this article.

\bigskip 
The article is structured as follows: section 2 briefly discusses
briefly the Schwarzschild spacetime in the so-called F-gauge in which
the construction of the cylinder at spatial infinity is carried out.
Section 3 is concerned with the Maxwell equations in the present
context. Section 4 is considers further properties of the
transport equations implied by the Maxwell equations at the cylinder
at spatial infinity, including some regularity conditions. Sections 5,
6 and 7 discuss, respectively, the solutions to the order $p$, $p+1$
and $p+2$ transport equations. There are some concluding remarks in
section 8. Finally, there are three appendices including the
definitions of some spinorial objects, the first orders of the
asymptotic expansions of the Schwarzschild solution, and some
properties of Jacobi polynomials which are used in the analysis.

\section{The Schwarzschild spacetime in the F-gauge}
The Schwarzschild spacetime was discussed in reference \cite{Fri98a} as an
example of the implementation of the framework of the cylinder at
spatial infinity ---or \emph{F-gauge}. This representation will be
used to analyse the propagation of the Maxwell field near spatial
infinity. Some of the relevant features of this construction are
reviewed. 

The Schwarzschild metric in isotropic coordinates is given by
\begin{equation}
\tilde{g}=\left(\frac{1-m/2r}{1+m/2r}\right)^2\mbox{d}t^2 +\left(1+\frac{m}{2r} \right)^4 \big( \mbox{d}r^2 + r^2 \mbox{d}\sigma^2\big).
\end{equation}
Now, consider the time symmetric hypersurface $\tilde{\mathcal{S}}$.
Writing $\tilde{h}=\Omega^{-2}h$, one can introduce a conformal metric
and conformal factor given by
\begin{equation}
h=-\big( \mbox{d}\rho^2 + \rho^2 \mbox{d}\sigma^2 \big), \quad \Omega=\frac{\rho^2}{(1+\rho m/2)^2},
\end{equation}
where the radial coordinate $\rho=1/r$ has been introduced. Let $i_1$
and $i_2$ denote the infinities corresponding to the two asymptotic
ends of the hypersurface $\tilde{\mathcal{S}}$. Further, let
$\mathcal{S}=\tilde{\mathcal{S}}\cup\{i_1,i_2 \}$. Let $i=i_1$ denote
the infinity corresponding to the locus $\rho=0$. The discussion in
this article will be concerned with the domain of influence,
$J^+(\mathcal{B}_a(i))$ of a sufficiently small ball,
$\mathcal{B}_a(i)$, of radius $a$ based on $i$. The point $i$ can be
blown up to a 2-sphere, $\mathcal{S}^2$. With this idea in mind
introduce the set $\mathcal{C}_a=(\mathcal{B}_a(i)\setminus
{i})\cup \mathcal{S}^2$.

The use of a gauge based on \emph{conformal Gaussian coordinates} to
discuss the structure of $\mathcal{B}_a(i)$ leads to a conformal
factor for the portion of spacetime under consideration which is given by
\begin{equation}
\Theta = \frac{\Omega}{\kappa}(1-\tau^2\frac{\kappa^2}{\omega^2}), \label{cf:Theta}
\end{equation}
where
\begin{equation}
\omega=\frac{2\Omega}{\sqrt{|D^i\Omega D_i\Omega|}}=\rho(1+\rho m/2),
\end{equation}
and $\kappa>0$ is a smooth function such that
$\kappa=\tilde{\kappa}\rho$ with $\tilde{\kappa}(i)=1$. The function
$\kappa$ encodes the remaining conformal freedom in the setting. In what
follows, and in order to ease the calculations let 
\begin{equation}
\kappa=\rho. \label{kappa}
\end{equation}
The coordinate $\tau$ is an affine parameter of conformal geodesics whose
tangent at $\tau=0$ is parallel to the normal of $\mathcal{S}$. The
coordinate $\rho$ can be extended off $\mathcal{S}$ by requiring it to
be constant along the aforementioned conformal geodesics. ``Angular
coordinates'' can be extended in a similar fashion.

In view of the conformal factor (\ref{cf:Theta}) define the manifold
\begin{equation}
\mathcal{M}_{a,\kappa} =\left\{ (\tau,q)\;|\; q\in \mathcal{C}_a, \; -\frac{\omega}{\kappa}\leq \tau \leq \frac{\omega}{\kappa} \right\},
\end{equation}
such that $J^+(\mathcal{B}_a(i))\subset\mathcal{M}_{a,\kappa}$, 
and the following relevant subsets thereof:
\begin{subequations}
\begin{eqnarray}
&& \mathcal{I}=\left\{(\tau,q)\in \mathcal{M}_{a,\kappa} \;|\; \rho(q)=0, |\tau|<\frac{\omega}{\kappa} \right\}, \\
&& \mathcal{I}^\pm = \left\{(\tau,q)\in \mathcal{M}_{a,\kappa} \;|\; \rho(q)=0, \tau=\pm 1\right\}, \\
&& \mathcal{I}^0 =\left\{(\tau,q)\in \mathcal{M}_{a,\kappa} \;|\; \rho=0, \;\tau=0  \right\}, \\
&& \mathscr{I}^\pm =\left\{ (\tau,q)\in \mathcal{M}_{a,\kappa} \;|\; q\in \mathcal{B}_a(i)\setminus \mathcal{I}^0, \;\tau=\pm\frac{\omega}{\kappa} \right\}, 
\end{eqnarray}
\end{subequations}
denoting, respectively, the \emph{cylinder at spatial infinity}, the
\emph{critical sets} where spatial infinity touches null infinity, the
intersection of the cylinder at spatial infinity with the initial
hypersurface $\mathcal{S}$, and the two components of \emph{null infinity}.

It will be convenient to work with a space-spinor formalism ---see
\cite{Som80}. In order to write down the field equations, introduce a
null frame $c_{AA'}$ satisfying
$g(c_{AA'},c_{BB'})=\epsilon_{AB}\epsilon_{A'B'}$. Let $\tau_{AA'}$
---with normalisation $\tau^{AA'}\tau_{AA'}=2$--- be tangent to the
conformal geodesics of which $\tau$ is a parameter. The frame can be
split in the form 
\begin{equation}
c_{AA'}=\frac{1}{2}\tau_{AA'} \tau^{CC'}c_{CC'}-\updn{\tau}{B}{A'}c_{AB},
\end{equation}
with
\begin{equation}
\tau^{AA'}c_{AA'}=\sqrt{2}\partial_{\tau}, \quad c_{AB}=\dnup{\tau}{(A}{B'}c_{B)B'}.
\end{equation}
For the purpose of the present investigation, the following choice is made:
\begin{equation}
c_{00'}=\frac{1}{\sqrt{2}}\left( (1+c^0)\partial_\tau+c^1\partial_\rho\right), \quad c_{11'}=\frac{1}{\sqrt{2}}\left( (1-c^0)\partial_\tau -c^1\partial_\rho \right).
\end{equation}
The remaining vectors of the frame, $c_{01'}$ and $c_{10'}$ must then
be tangent to the spheres $\{\tau=\mbox{constant}, \;
\rho=\mbox{constant}\}$, and thus cannot define smooth vector fields
everywhere. To avoid this difficulty all possible tangent vectors
$c_{01'}$ and $c_{10'}$ will be considered. This results in a
5-dimensional submanifold of the bundle of normalised spin
frames. Rotations $c_{01'}\rightarrow \mbox{e}^{\mbox{i}\alpha}$,
$\alpha\in\Real$ leave this submanifold invariant. Hence it defines a
subbundle with structure group $U(1)$ which projects into
$\mathcal{M}_{a,\kappa}$. All the relevant structures will be lifted
to the subbundle, which in an abuse of notation will be again denoted
by $\mathcal{M}_{a,\kappa}$.

The introduction of the bundle space $\mathcal{M}_{a,\kappa}$ and of
the frame $c_{AA'}$ in our formalism implies that all relevant
quantities have a definite spin weight and hence admit a
characteristic ---spherical harmonics--- expansion in terms of some
functions $\TT{j}{k}{l}$ associated with unitary representations of
$SU(2,\Complex)$ ---see e.g. \cite{Fri98a,Fri04} for a more detailed
discussion in this respect. Further, one can introduce differential
operators $X$, $X_+$ and $X_-$ defined by their action on the
functions $\TT{j}{k}{l}$. It is noted that with the help of these operators one can write
\begin{equation}
c_{AA'}=c^\mu_{AA'}\partial_\mu=c^0_{AA'}\partial_\tau + c^1_{AA'}\partial_\rho+ c^+_{AA'}X_+ +c^-_{AA'}X_-.
\end{equation}

In addition to the frame $c_{AA'}$, the geometry of
$\mathcal{M}_{a,\kappa}$ is described by means of the associated
connection $\Gamma_{AA'BC}$, the spinorial counterpart of the Ricci
tensor of a Weyl connection $\Theta_{AA'BB'}$ and the rescaled Weyl
spinor $\phi_{ABCD}$. The unprimed (space-spinor) version of the
connection spinor is given by
$\Gamma_{ABCD}=\dnup{\tau}{B}{B'}\Gamma_{AB'CD}$,
$\Gamma_{AA'CD}=\Gamma_{ABCD}\updn{\tau}{B}{A'}$, which is decomposed
as
\begin{equation}
\Gamma_{ABCD}=\frac{1}{\sqrt{2}}\left(\xi_{ABCD}-\chi_{(AB)CD} \right)-\frac{1}{2}\epsilon_{AB}f_{CD}.
\end{equation}
Similarly, one considers
$\Theta_{ABCD}=\dnup{\tau}{C}{A'}\dnup{\tau}{D}{B'}\Theta_{AA'BB'}$.
The explicit spherical symmetry of the spacetime justifies the
following Ansatz in terms of irreducible spinors:
\begin{subequations}
\begin{eqnarray}
&& c^0_{AB}=c^0 x_{AB}, \quad c^1_{AB}=c^1x_{AB}, \quad c^-_{AB}=c^-y_{AB}, \quad c^+_{AB}=c^+z_{AB}, \label{unknown_1}\\
&& f_{AB}=f x_{AB}, \quad \xi_{ABCD}=\xi (\epsilon_{AC}x_{BD} +\epsilon_{BD}x_{AC}),  \label{unknown_2}\\
&& \chi_{(AB)CD}=\chi_2 \epsilon^2_{ABCD} + \chi_h h_{ABCD}, \label{unknown_3}\\
&& \Theta_{ABCD}=\Theta_2 \epsilon^2_{ABCD}+\Theta_h h_{ABCD}+\Theta_x \epsilon_{AB}x_{CD}, \quad \phi_{ABCD}=\phi\epsilon^2_{ABCD}. \label{unknown_4}
\end{eqnarray}
\end{subequations}
The definitions of the irreducible spinors introduced above are given
in the appendix. The manifest spherical symmetry of this
representation implies that the functions $c^0$, $c^1$, $c^\pm$, $f$,
$\xi$, $\chi_2$, $\chi_h$, $\phi$ have spin-weight 0. Furthermore they
only contain the function $\TT{0}{0}{0}=1$.

The functions $c^0$, $c^1$, $c^\pm$, $f$, $\xi$, $\chi_2$, $\chi_h$,
$\phi$ are determined by solving the conformal propagation equations
discussed in \cite{Fri98a} with the appropriate initial data. The
problem of reconstructing the conformal Schwarzschild solution from
the given data amounts to finding a solution $u=u(\tau,\rho;m)$ of an
initial value problem of the type
\begin{equation}
\partial_\rho u =F(u,\tau,\rho;m), \quad u(0,\rho;m)=u_0(\rho;m), \label{system:Schwarzschild}
\end{equation}
with analytic functions $F$ and $u_0$. The solution with $m=0$
corresponds to a portion of the conformal Minkowski spacetime, in
which the only non-vanishing components of the solution are given by
\begin{equation}
c^0=-\tau, \quad c^1=\rho, \quad c^\pm=1, \quad f=1.
\end{equation}
Since in this case the solution exists for all $\tau$, $\rho\in
\Real$, it follows from standard results of ordinary differential
equations that for a given $m$ there is sufficiently small $\rho_0$
such that there is an analytic solution to the system
(\ref{system:Schwarzschild}) which extends beyond $\scri$ for $|\rho|
< |\rho_0|$. Hence, one can recover the portion of the Schwarzschild
spacetime which lies near null and spatial infinity,
$J^+(\mathcal{B}_a(i))$, if $a$ is taken to be small enough.

It follows from the above discussion that the coefficients that are
obtained from solving the transport propagation equations on the
cylinder at spatial infinity ---as discussed in
\cite{Fri98a,Val04a}--- correspond to the first terms in the
expansions of the solutions of (\ref{system:Schwarzschild}). See the
appendix for a list of the relevant expansions.

\section{The Maxwell field}
As it is common when working with spinors, consider the Maxwell tensor
written as
\begin{equation}
F_{AA'BB'}=\phi_{AB}\epsilon_{A'B'}+\bar{\phi}_{A'B'}\epsilon_{AB}.
\end{equation}
Then the Maxwell equations are equivalent to
\begin{equation}
\nabla^{AA'}\phi_{AB}=0,
\end{equation}
that is, \emph{the spin-1 zero-rest-mass field equations}. Recall, as
well, that if the conformal weight of $\phi_{AB}$ is chosen properly,
the vacuum Maxwell equations are conformally invariant. The spinor
$\phi_{AB}$ is symmetric and thus, one can write
\begin{equation}
\phi_{AB}=\phi_0\epsilon^0_{AB} +\phi_1\epsilon^1_{AB}+\phi_2 \epsilon^2_{AB}.
\end{equation}

In the F-gauge, and using a space-spinor decomposition, the above equation is equivalent to the following propagation equations:
\begin{subequations}
\begin{eqnarray}
&& (\sqrt{2}-2c^0_{01})\partial_\tau \phi_0 +2c^0_{00}\partial_\tau \phi_1 -2c^\alpha_{01}\partial_\alpha\phi_0+2c^\alpha_{00}\partial_\alpha\phi_1 \nonumber \\
&&\hspace{2cm}=(2\Gamma_{0011}-4\Gamma_{1010})\phi_0+4\Gamma_{1000}\phi_1-2\Gamma_{0000}\phi_2, \label{p0}\\
&& \sqrt{2}\partial_\tau \phi_1 -c^0_{11}\partial_\tau\phi_0+c^0_{00}\partial_\tau\phi_2-c^\alpha_{11}\partial_\alpha\phi_0+c^\alpha_{00}\partial_\alpha\phi_2 \nonumber \\
&&\hspace{2cm}=-(2\Gamma_{1110}+f_{11})\phi_0 +2(\Gamma_{1100}+\Gamma_{0011})\phi_1-(2\Gamma_{0001}-f_{00})\phi_2, \label{p1}\\
&& (\sqrt{2}+2c^0_{01})\partial_\tau \phi_2 -2c^0_{11}\partial_\tau\phi_1 + 2c^\alpha_{01}\partial_\alpha \phi_2-2c^\alpha_{11}\partial_\alpha\phi_1 \nonumber \\
&&\hspace{2cm}=-2\Gamma_{1111}\phi_0+4\Gamma_{0111}\phi_1+(2\Gamma_{1100}-4\Gamma_{0101})\phi_2. \label{p2}
\end{eqnarray}
\end{subequations}
In addition, one has the following constraint equation
\begin{eqnarray}
&&c^0_{11}\partial_\tau \phi_0-2c^0_{01}\partial_\tau\phi_1 +c^0_{00}\partial_\tau\phi_2 +c^\alpha_{11}\partial_\alpha \phi_0-2c^\alpha_{01}\partial_\alpha\phi_1+c^\alpha_{00}\partial_\alpha\phi_2 \nonumber \\
&&\hspace{2cm}= (2\Gamma_{1110}-2\Gamma_{(01)11})\phi_0+2(\Gamma_{0011}-\Gamma_{1100})\phi_1 +(2\Gamma_{(01)00}-\Gamma_{0001})\phi_2. \label{c1}
\end{eqnarray}
Because of the symmetries of the Schwarzschild spacetime, the
connection coefficients $\Gamma_{0011}$, $\Gamma_{1010}$,
$\Gamma_{1100}$ and $\Gamma_{0101}$ are the only non-vanishing ones.
The leading terms of frame and connection components are listed in the
appendix \ref{appendix:expansions}.

Writing $\phi=(\phi_0,\phi_1,\phi_2)$, the Maxwell propagation equations
can be concisely expressed in the form
\begin{equation} \label{maxwell:propagation}
\sqrt{2}E\partial_\tau + A^{AB}c^\mu_{AB}\partial_\mu \phi =B(\Gamma_{ABCD})\phi,
\end{equation}
where $E$ denotes the $(3\times 3)$-unit matrix, $A^{AB}c^\mu_{AB}$ are $(3\times 3)$-matrices depending on the
coordinates and $B(\Gamma)$ is a matrix valued function of the connection $\Gamma_{ABCD}$. For consistency
the constraint equation (\ref{c1}) will be written as
\begin{equation} \label{maxwell:constraint} F^{AB}c^\mu_{AB}
  \partial_\mu\phi=H(\Gamma_{ABCD})\phi,
\end{equation}
where $F^{AB}c^\mu_{AB}$ denotes $(1\times 3)$ matrices, and $H$ is a
function of the connection.

The system (\ref{maxwell:propagation}) shares the property with the Bianchi
propagation system for the rescaled Weyl tensor and the propagation
system for linearised gravity that its principal
degenerates at critical sets $\mathcal{I}^\pm$. More precisely, one
has that
\begin{equation}
(\sqrt{2}E +A^{AB}c^0_{AB})|_{\mathcal{I}}=\sqrt{2}\mbox{diag}(1-\tau,1,1+\tau).
\end{equation}
Most non-smooth and/or polyhomogeneous behaviour of the solutions of
the Maxwell equations at the conformal boundary can be traced back to
precisely this degeneracy.

Another deciding property of the Maxwell equations in the
aforediscussed form stems from the fact that
$A^{AB}c^1_{AB}|_{\mathcal{I}}=0$. From here it follows ---see e.g.
the discussion in \cite{Val03a}--- that the equations imply an
intrinsic system of transport equations on $\mathcal{I}$ which
determines the solution on $\mathcal{I}$ in terms of the data implied
by $\phi$ on $\mathcal{I}^0$. In other words, $\mathcal{I}$ is a
\emph{total characteristic} of (\ref{maxwell:propagation}) and
(\ref{maxwell:constraint}).

As in the case of the conformal Einstein field equations of
\cite{Fri98a}, given an asymptotically flat spacetime one can produce,
for the Maxwell field, a hierarchy of transport equations at the
cylinder at spatial infinity out of the equations
(\ref{maxwell:propagation}) and (\ref{maxwell:constraint}).
Differentiating the latter equations with respect to $\rho$, say, $p$
times and evaluating them at the cylinder at spatial infinity $\mathcal{I}$
one obtains, by virtue of $c^1_{AB}|_{\mathcal{I}}=0$ the equations
\begin{subequations}
\begin{eqnarray}
&& (\sqrt{2}E + A^{AB}(c^0_{AB})^{(0)}) \partial_\tau \phi^{(p)} + A^{AB}(c^C_{AB})^{(0)}\partial_C \phi^{(p)} \nonumber \\
&&\hspace{1.2cm}=B(\Gamma^{(0)}_{ABCD})\phi^{(p)} +\sum_{j=1}^p \binom{p}{j} ( B(\Gamma_{ABCD})\phi^{(p-j)}-A^{AB}(c^\mu_{AB})^{(j)}\partial_\mu \phi^{(p-j)}), \label{transport_1}\\
&& F^{AB}(c^0_{AB})^{(0)}\partial_\tau \phi^{(p)} + F^{AB}(c^C_{AB})^{(0)}\partial_C\phi^{(p)} \nonumber \\
&&\hspace{1.2cm}=H(\Gamma^{(0)}_{ABCD})\phi^{(p)} +\sum_{j=1}^p \binom{p}{j} ( H(\Gamma_{ABCD})\phi^{(p-j)}-F^{AB}(c^\mu_{AB})^{(j)}\partial_\mu \phi^{(p-j)}), \label{transport_2}
\end{eqnarray}
\end{subequations}
with $C=\pm$. It is noted that the principal part of the above
equations is universal, in the sense that it depends only on the
zeroth-order solution of the background metric. The first term on the
right-hand side is also universal, while the remaining non-homogeneous
terms depend on lower order terms of the Maxwell field and higher
order terms of the background spacetime.  As in the case of the
gravitational field, provided some suitable initial data one can solve
the above equations to any desired order in a recursive way.

The coefficients $\phi_0$, $\phi_1$ and $\phi_2$ have, respectively,
spin-weights $1$, $0$ and $-1$. Consistently with their spin-weight,
the following Ansatz will be made:
\begin{equation}
  \phi_j =\sum_{p=|1-j|}^\infty \sum_{q=|1-j|}^p \sum_{k=0}^{2q} \frac{1}{p!} a_{j,p;q,k}(\tau) \TT{2q}{k}{q-1+j}\rho^p, \label{Ansatz:expansion}
\end{equation}
with $j=0,1,2$, and $a_{j,p;2q,k}: \Real \longrightarrow \Complex$. 

The above assumption will allow to solve the hierarchy of transport
equations (\ref{transport_1}) and (\ref{transport_2}) by ``decomposing
in modes''. It is important to note that because the background
terms in the transport equations only contain the function
$\TT{0}{0}{0}=1$, it will not be necessary to linearise products of
the form $\TT{j_1}{l_1}{m_1}\times \TT{j_2}{l_2}{m_2}$ when performing
the mode decomposition.

\subsection{Initial data for the Maxwell equations}
The equation (\ref{c1}) evaluated on $\tau=0$ reduces, for the
Schwarzschild spacetime, to
\begin{equation}
\rho \partial_\rho \phi_1 +\frac{1}{2} X_-\phi_0 -\frac{1}{2} X_+\phi_2=0.
\end{equation}
The latter equation can be read as an equation (ODE) for $\phi_1$, if
$\phi_0$ and $\phi_2$ are given on the initial hypersurface. In the
present work, initial data for the Maxwell equations on the initial
hypersurface $\mathcal{S}$ which when lifted to $\mathcal{C}_a$ is expandable in
powers of $\rho$ will be considered. Accordingly, prescribe on
$\mathcal{C}_a$ components $\phi_0$ and $\phi_2$ of the form
\begin{equation}
\phi_0=\sum_{p=0}^\infty \sum_{q=1}^p \sum_{k=0}^{2q} \frac{1}{p!}a_{0,p;2q,k}(0) \TT{q}{k}{q-1}\rho^p, \quad  \phi_2=\sum_{p=0}^\infty \sum_{q=1}^p \sum_{k=0}^{2q} \frac{1}{p!}a_{2,p;q,k}(0) \TT{2q}{k}{q+1}\rho^p,
\end{equation}
which correspond to the expansions of analytic $\phi_0$ and $\phi_2$
---see \cite{Fri98a}. One can use equation (\ref{c1}) to obtain the
coefficients $a_{1,p;2q,k}(0)$ in
\begin{equation}
\phi_1=\sum_{p=0}^\infty \sum_{q=0}^p \sum_{k=0}^{2q} \frac{1}{p!} a_{1,p;q,k}(0) \TT{2q}{k}{q}\rho^p,
\end{equation}
in terms of $a_{0,p;2q,k}(0)$ and $a_{2,p;2q,k}(0)$. It is noted that in
this construction the coefficient $a_{1,0;0,0}(0)$ ---the electric
charge--- is not determined and can be prescribed freely. On the other
hand, from equation (\ref{c1}) it follows that $a_{1,p;0,0}(0)=0$ for $p\geq 1$.

\section{Further properties of the transport equations}
The discussion in this section is inspired from the treatment given in
$\cite{Fri98a}$ of the transport equations for the Conformal Einstein
equations.  Besides being given as a reference, it will help to
highlight the difficulties to be confronted when analysing the
solutions of the equations (\ref{transport_1}) and (\ref{transport_2})
for an arbitrary order $p$.

\subsection{Reduction to an overdetermined system of ODE's}
The propagation transport equations implied by the Maxwell equations on the cylinder at spatial infinity of the Schwarzschild background are of the form:
\begin{subequations}
\begin{eqnarray}
&& (1+\tau)\partial_\tau \phi_0^{(p)}+X_+\phi_1^{(p)}-(p-1)\phi_0^{(p)}=R_0^{(p)}, \\
&& \partial_\tau \phi_1^{(p)}+\frac{1}{2}(X_+\phi_2^{(p)}+X_-\phi_0^{(p)})=R_1^{(p)}, \\
&& (1-\tau)\partial_\tau \phi_2^{(p)} + X_-\phi_1^{(p)}+(p-1)\phi_2^{(p)}=R_2^{(p)},
\end{eqnarray}
\end{subequations} 
while the constraint transport equation is given by
\begin{equation}
\tau\partial_\tau\phi_1^{(p)}+\frac{1}{2}X_+\phi_2^{(p)}-\frac{1}{2}X_-\phi_0^{(p)}-p\phi_1^{(p)}=S^{(p)}.
\end{equation}
As mentioned before, the terms $R_1^{(p)}$, $R_2^{(p)}$, $R_3^{(p)}$
and $S^{(p)}$ depend solely on the lowers order solutions to the
equations. Using the expansion Ansatz (\ref{Ansatz:expansion}) it can be seen
that the coefficients $a_{j,p;q,k}$ satisfy the equations
\begin{subequations}
\begin{eqnarray}
&& (1+\tau) a_{0,p;q,k}' +\sqrt{q(q+1)} a_{1,p;q,k}-(p-1)a_{0,p;q,k}= U_{0,p;q,k}, \label{tp0}\\
&& a_{1,p;q,k}'+\frac{1}{2}\sqrt{q(q+1)}(a_{2,p;q,k}-a_{0,p;q,k})=U_{1,p;q,k}, \label{tp1}\\
&& (1-\tau)a_{2,p;q,k}'-\sqrt{q(q+1)}a_{1,p;q,k}+(p-1)a_{2,p;q,k}=U_{2,p;q,k}, \label{tp2}
\end{eqnarray}
\end{subequations}
and
\begin{equation}
\tau a_{1,p;q,k}'+\frac{1}{2}\sqrt{q(q+1)}(a_{2,p;q,k}+a_{0,p;q,k})-pa_{1,p;q,k}=T_{p;q,k}, \label{tc1}
\end{equation}
for $p\geq 1$, $1\leq q \leq p$, and where again the nonhomogeneous
terms $U_{0,p;q,k}$, $U_{1,p;q,k}$, $U_{2,p;q,k}$ and $T_{p;q,k}$ in
the equations depend on the background spacetime and lower order
$a_{j,p;q,k}$'s. The problem has been reduced to the analysis of an
overdetermined system of ordinary differential equations. The last
equation can be used to eliminate the unknown $a_{1,p;q,k}$. In what
remains of this section, attention will be focused on a fixed ---but
otherwise arbitrary--- value of the indices $p$, $q$, $k$ in their
range of definition which henceforth will often be suppressed in order
to avoid cluttering the equations.

The coefficient $a_1$ can be eliminated from the following
considerations by noting that
\begin{equation}
-\frac{1}{2}\sqrt{q(q+1)}(1+\tau)a_0-\frac{1}{2}\sqrt{q(q+1)}(1-\tau)a_2+pa_1=\tau U_1-T, \label{auxiliary}
\end{equation}
from where one can solve for $a_1$. One is left with the system
\begin{subequations}
\begin{eqnarray}
&& (1+\tau)a'_0+\left(\frac{1}{2}q(q+1)(1+\tau)-p+1  \right)a_0 +\frac{1}{2p}q(q+1)(1-\tau)a_2 \nonumber \\
&&\hspace{7cm}=U_0+\frac{1}{p}\sqrt{q(q+1)}(T-\tau U_1), \label{new_tp0} \\
&& (1-\tau)a'_2-\frac{1}{2p}q(q+1)(1+\tau)a_0+\left(-\frac{1}{2p}q(q+1)(1-\tau)+p-1  \right)a_2 \nonumber \\
&&\hspace{7cm}=U_2+\frac{1}{p}\sqrt{q(q+1)}(\tau U_1-T). \label{new_tp2}
\end{eqnarray}
\end{subequations}
For future reference the above system will be written in matricial form as
\begin{equation}
Y_{p;q,k}'(\tau)=A_{p;q,k}(\tau)Y_{p;q,k}(\tau)+B_{p;q,k}(\tau),
\end{equation}
with $A_{p;q,k}$ a $(2\times 2)$-matrix, and $Y_{p;q,k}$,
$B_{p;q,k}$ $(2\times 1)$-column vectors. Note that the Wronskian determinant 
$\det X_{p;q}$ of the above system ---satisfying $(\det
X_{p;q})'=(\mbox{Tr}A_{p;q,k})X_{p;q}$--- is given by
\begin{equation}
\det X_{p;q}=X^0_{p;q}(1-\tau^2)^{p-1}, \label{Wronskian}
\end{equation}
where $X^0_{p;q}$ is a constant. In order to calculate the fundamental
matrix of the system consider the equations (again, the multisubindex begin supressed)
\begin{subequations}
\begin{eqnarray}
&& (1-\tau^2)a_0'' + \big( 2-(2-2p)\tau\big)a_0' +(q+p)(q-p+1)=0, \label{Jacobi_1}\\
&& (1-\tau^2)a_2'' + \big( -2-(2-2p)\tau \big)a_2' +(q+p)(q-p+1)=0, \label{Jacobi_2}
\end{eqnarray}
\end{subequations}
which are implied by the homogeneous parts of the equations
(\ref{new_tp0}) and (\ref{new_tp2}). Equations (\ref{Jacobi_1}) and (\ref{Jacobi_2}) are so-called Jacobi equations
\begin{equation}
D_{(n,\alpha,\beta)}a=(1-\tau)a''+\big(\beta-\alpha-(\alpha+\beta+2)\tau\big)a' +n(n+\alpha+\beta+1)a=0,
\end{equation}
with
\begin{equation}
\alpha=-1-p, \quad \beta=1-p, \quad n_1=p+q, \quad n_2=p-q-1.
\end{equation}
Note that if $a_0(\tau)$ is a solution of equation (\ref{Jacobi_1}),
then $a^s_{0}(\tau)\equiv a_0(-\tau)$ is a solution of equation
(\ref{Jacobi_2}). A class of solutions is given by the Jacobi
polynomials $P^{(\alpha,\beta)}_n(\tau)$ ---see the appendix for their
definition and relevant properties.

Using the definition of the polynomials, $P_{p+q}^{(-1-p,-p+1)}$ are
seen to vanish identically. On the other hand,
\begin{equation}
Q_2=P_{p+q}^{(-p-1,-p+1)}
\end{equation}
is a solution of degree $n_2=p-q-1$ for $p\geq 2$, $ 1\leq q\leq p-1$.
Using some identities of the polynomials one finds that
\begin{equation}
Q_1=\left( \frac{1-\tau}{2} \right)^{p+1} P^{(p+1,-p+1)}_{q-1},
\end{equation}
is another solution, $p\geq 2$, $1\leq q \leq p-1$. Clearly $Q_1$ and
$Q_2$ are linearly independent as $n_1>n_2$. Another solution is given by
\begin{equation}
Q_3=\left(\frac{1+\tau}{2}\right)^{p-1}P_{q+1}^{(-p-1,p-1)},
\end{equation}
for $p\geq 2$, $1\leq q \leq p-1$. A direct calculation shows that the
fundamental matrix, $X_{p,q}$ of the system is given by
\begin{equation}
X_{p,q}=\begin{pmatrix}
  Q_1 & (-1)^{q+1} Q_3 \\
  (-1)^{q+1} Q^s_{3} & Q^s_{1}
  \end{pmatrix}
\end{equation}
for $p\geq 2$, $1\leq q \leq p-1$. The notation $Q^s_1(\tau) \equiv
Q_1(-\tau)$, $Q^s_3(\tau)\equiv Q_3(-\tau)$ has been used again. Note that
the fundamental matrix does not depend on $k$. The inverse of
$X_{p,q}$ is given by
\begin{equation}
X_{p,q}^{-1}=\frac{1}{X^0_{p;q}}\begin{pmatrix}
  \frac{\displaystyle (1+\tau)^2}{\displaystyle (1-\tau)^{p-1}}P_{q-1}^{(p+1,1-p)}(-\tau) & \frac{\displaystyle (-1)^{q}}{\displaystyle (1-\tau)^{p-1}}P_{q+1}^{(-p-1,p-1)}(\tau) \\
  \frac{\displaystyle (-1)^{q}}{\displaystyle (1+\tau)^{p-1}}P_{q+1}^{(-p-1,p-1)}(-\tau) & \frac{\displaystyle (1-\tau)^2}{\displaystyle (1+\tau)^{p-1}}P_{q-1}^{(p+1,1-p)}(\tau)
  \end{pmatrix}.
\end{equation}

The case $p=q$, $p \geq 1$ is solved separately. It is noted that
\begin{equation}
D_{(2p,-p-1,-p+1)}\left( \left(\frac{1-\tau}{2}\right)^{p+1} \left( \frac{1+\tau}{2}\right)^{p-1} a_0 \right)= \left( \frac{1-\tau}{2}\right)^{p+1} \left(\frac{1+\tau}{2}\right)^{p-1} D_{(0,p+1,p-1)}a_0, 
\end{equation}
where
\begin{equation}
D_{(0,p+1,p-1)}a_0=(1-\tau^2) a''_0 +(-2-(2p+2))a'_0,
\end{equation}
so that
\begin{equation}
a_0=\left( \frac{1-\tau}{2}\right)^{p+1} \left(\frac{1+\tau}{2}\right)^{p-1}\left(C_1+C_2 \int_0^\tau \frac{\mbox{d}s}{(1+s)^p(1-s)^{p+2}}\right),
\end{equation}
with $C_1$ and $C_2$ some integration constants. Noting, again, that
$a_2(\tau)=a^s_{0}(\tau)\equiv a_0(-\tau)$ one obtains the following

\begin{proposition}
  For $p=q$, the fundamental matrix $X_{p;q}$ of the system
  (\ref{new_tp0}) and (\ref{new_tp2}) is polynomial if and only if
\begin{equation}
a_{0,p;q,k}(0)=a_{2,p;q,k}(0). \label{regularity:condition}
\end{equation}  
\end{proposition}
The latter is the analogue, in our context, of the regularity
conditions of \cite{Fri98a} and \cite{Val03a}. 

\bigskip
\emph{In what
  follows, it will be assumed that the initial data for the propagation
  equations (\ref{p0})--(\ref{p2}) is constructed such that 
  condition (\ref{regularity:condition}) holds. }

\subsection{The solution to the nonhomogenous part}
Once the fundamental matrix of the system
(\ref{new_tp0})--(\ref{new_tp2}) is known, the solution is given by
\begin{equation}
\binom{a_{0,p;q,k}(\tau)}{a_{2,p;q,k}(\tau)}=X_{p,q}(\tau) \left[ X^{-1}_{p,q}(0)\binom{a_{0,p;q,k}(0)}{a_{2,p;q,k}(0)} +\int^\tau_0 X^{-1}_{p,q}(s)B_{p,q}(s) \mbox{d}s \right]. \label{solution:integral}
\end{equation}
As seen before, the entries of $X^{-1}_{p;q}$ are rational functions with
---in the worst of cases--- poles of order $p-1$ at $\tau=\pm 1$. One
can make the following direct but crucial statement:

\begin{proposition}
  If the functions $U_{0,p;q,k}$, $U_{1,p;q,k}$, $U_{2,p;q,k}$ and
  $T_{p;q,k}$ in equations (\ref{p0})--(\ref{p2}) and (\ref{c1}) are
  polynomial in $\tau$ ---that is, if the lower order functions
  $a_{j,p';q',k'}(\tau)$ with $0\leq p'< p$, $0\geq q'\geq p'$, $0\geq k'\geq 2q'$ are polynomials in $\tau$--- then the
  solutions given by (\ref{solution:integral}) consist of polynomials
  in $\tau$ and of products of polynomials in $\tau$ times the functions
  $\ln(1-\tau)$ and $\ln(1+\tau)$.
\end{proposition}

\textbf{Proof.} The entries of the vector $X^{-1}_{p,q}B_{p,q}$ are, respectively, of the form
\begin{equation}
  \frac{f_0(\tau)}{(1+\tau)(1-\tau)^{p-1}} \quad \mbox{and} \quad \frac{f_2(\tau)}{(1-\tau)(1+\tau)^{p-1}},
\end{equation}
where $f_0(\tau)$ and $f_2(\tau)$ are polynomials in $\tau$.
Decomposition in \emph{partial fractions} of the first entry renders
an expression containing $(1+\tau)^{-1}$, $(1-\tau)^{-n}$ with
$n=1,\ldots,p-1$, and a polynomial. Analogously for the second entry,
where one finds a decomposition in terms of $(1-\tau)^{-1}$,
$(1+\tau)^{-n}$ with $n=1,\ldots,p-1$, and a polynomial. The integration
produces, in the first entry, an expression containing
$\ln(1\pm\tau)$, $(1-\tau)^{1-n}$ with $n=2,\ldots,p-1$ and a
polynomial. Similarly for the second entry. Now, the multiplication by
the fundamental matrix cancels out the rational expressions. \hfill $\blacksquare$

\bigskip
\textbf{Remark 1.}  
From the previous discussion it is clear that even if the
regularity condition (\ref{regularity:condition}) holds, the solutions
to the transport equations will contain $\ln(1\pm\tau)$ terms unless
some cancellations take place.

\textbf{Remark 2.} On the other hand, one could use an argument
analogous to that of the above proof to show that if the lower order
coefficients already contain $\ln(1\pm\tau)$-singular terms ---say
because the regularity condition (\ref{regularity:condition}) is not
satisfied, then the integration in equation (\ref{solution:integral})
would render $\ln^2(1\pm\tau)$-terms. This type of analysis is, however,
beyond the present discussion.

\section{Expansions at order $p$}

From the hyperbolicity and linearity of the propagation equations, if
for a given order $p$ a certain type of harmonics has not yet appeared
in the initial data, it will also not be present in the solutions to
the transport equations. For the class of initial data under
consideration, terms containing harmonics of the form
$\TT{2p}{k}{p-1+j}$ ---which will be referred to as $2^p$-polar---
appear firstly at order $p$. It follows that if $p=q$, then
\begin{equation}
B_{p;p,k}(\tau)=0, 
\end{equation}
for the class of data under consideration ---for all the lower terms
$a_{j,p';q',k'}$ with $0\leq p'\leq p$, $|1-j| \leq q'\leq p'$ and
$0\leq k' \leq 2q'$ can only contain at most $\TT{2(p-1)}{k}{p-2+j}$
functions. In the sequel, the following observation will be repeatedly
used: the Schwarzschild solution contains only $\TT{0}{0}{0}$ sectors.
Hence, all the products of coefficients of the Maxwell field and
background terms are trivially expanded in terms of the functions
$\TT{2q}{k}{p-2+j}$ present in the coefficients of the Maxwell field.
This observation might seem trivial, but note that if one considers a
representation of the Schwarzschild solution where the spherical
symmetry is not manifest, a ``mixing of modes'' will occur ---and in
particular higher order harmonics would be generated, severely
complicating the analysis. From here it follows that the first
appearance in the solutions to the transport equations of terms
containing $\TT{2p}{k}{p-1+j}$ renders no logarithms if the regularity
condition (\ref{regularity:condition}) is satisfied.

\begin{proposition} \label{proposition:p}
  If the regularity condition (\ref{regularity:condition}) holds, then
  the solutions $a_{0,p;p,k}(\tau)$, $a_{1,p;p,k}(\tau)$ and
  $a_{2,p;p,k}(\tau)$ of the transport equations
  (\ref{tp0})--(\ref{tp2}) and (\ref{tc1}) for $p=q$ and
  $k=0,\ldots,2p$ are polynomial in $\tau$. Moreover, one has that
\begin{subequations}
\begin{eqnarray}
&& a_{0,p;p,k}(\tau)= C_{p;p,k} (1-\tau)^{p+1}(1+\tau)^{p-1}, \label{a0p} \\
&& a_{1,p;p,k}(\tau)= C_{p;p,k}\sqrt{\frac{p+1}{p}}(1-\tau)^p(1+\tau)^p, \label{a1p} \\
&& a_{2,p;p,k}(\tau)= C_{p;p,k} (1-\tau)^{p-1}(1+\tau)^{p+1}, \label{a2p}
\end{eqnarray}
\end{subequations}
with $C_{p;p,k}=a_{0,p;p,k}(0)$.
\end{proposition}

\section{Expansions at order $p+1$}

If the condition (\ref{regularity:condition}) holds, then the first
occasion where the harmonics $\TT{2p}{k}{p-1+j}$ can produce
logarithms via the formula (\ref{solution:integral}) is in the order
$p+1$ of the transport equations. Quite remarkably, as it will be
seen, this is not the case: cancellations occur which prevent the
appearance of logarithms at this order. It is necessary to go up to
order $p+2$ to find the first logarithms in the sectors
$\TT{2p}{k}{p-1+j}$. More precisely, one has the following:

\begin{proposition} \label{proposition:p+1}
If the regularity condition (\ref{regularity:condition}) is satisfied for
$p\in \Natural$, $p\geq 1$ then the functions $a_{j,p;2p,k}(\tau)$ are
polynomial in $\tau$ ---that is, they are logarithm free.
\end{proposition}

\textbf{Remark.} It must be emphasised that in view of the formula
(\ref{solution:integral}), the above result arises as the result of a
remarkable collusion of terms which, in principle, there is no reason
to expect.  It would be very interesting to understand the deeper
(possibly group theoretical) reasons behind this result.

\textbf{Proof.} The proof proceeds by an explicit calculation. The equations
(\ref{new_tp0})--(\ref{new_tp2}) imply upon the replacement $p\rightarrow
p+1$, $q\rightarrow p$ the system
\begin{subequations}
\begin{eqnarray}
&& a'_{0,p+1} -
\frac{p}{2}\frac{(1-\tau)}{(1+\tau)}a_{0,p+1}+\frac{p}{2}\frac{(1-\tau)}{(1+\tau)}a_{2,p+1}=b_{0,p+1},
\\
&& a'_{2,p+1}+\frac{p}{2}\frac{(1+\tau)}{(1-\tau)}a_{2,p+1}-\frac{p}{2}\frac{(1+\tau)}{(1-\tau)}a_{0,p+1}=b_{2,p+1},
\end{eqnarray}
\end{subequations}
where the subindex string ${}_{;p,k}$ has been suppressed throughout for the
ease of the presentation. In accordance with equation (\ref{transport_1}) one
has
\begin{subequations}
\begin{eqnarray}
&& b_{0,p+1}= \frac{1}{1+\tau}\big(h_{0,p} a_{0,p}+h_{1,p}a_{1,p} + h_{2,p} a_{2,p} + \tilde{h}_{0,p}
a'_{0,p}+\tilde{h}_{1,p} a'_{1,p}+\tilde{h}_{2,p} a'_{2,p} \big), \\
&& b_{2,p+1}= \frac{1}{1-\tau}\big(h_{2,p}^s a_{0,p}+h_{1,p}^sa_{1,p} + h_{0,p}^s a_{2,p} + \tilde{h}_{2,p}^s
a'_{0,p}+\tilde{h}_{1,p}^s a'_{1,p}+\tilde{h}_{0,p}^s a'_{2,p} \big),
\end{eqnarray}
\end{subequations}
with $h_{j,p}$ and $\tilde{h}_{j,p}$ for $j=0,1,2$ some polynomials in
$\tau$ arising from the order $0,1,2$ solutions of the transport
equations for the Schwarzschild background ---see appendix
\ref{appendix:expansions}. As before the notation
$h_{j,p}^s(\tau)\equiv h_{j,p}(-\tau)$ has been used.

The order $p$ Maxwell transport equations can be used to eliminate $a'_{0,p}$,
$a'_{1,p}$ and $a'_{2,p}$ from the expressions for $b_{0,p+1}$ and
$b_{2,p+1}$. Further, equation (\ref{auxiliary}) with $q=p$ can
be used to eliminate $a_{1,p+1}$. At the end one is left with
\begin{subequations}
\begin{eqnarray}
&& b_{0,p+1}=\frac{1}{(1+\tau)^2}\big( f_0 a_{0,p} + f_2 a_{2,p} \big),
\\
&& b_{2,p+1}=\frac{1}{(1-\tau)^2}\big( f_{2}^s a_{0,p} + f^s_{0} a_{2,p} \big),
\end{eqnarray}
\end{subequations}
with 
\begin{subequations}
\begin{eqnarray}
&& f_0(\tau)=-\frac{1}{12}m(p+1)\tau \big( (-4p-5)\tau^6 +(-8p-2)\tau^5 +(25+16p)\tau^4+(56p+40)\tau^3 \nonumber \\
&&\hspace{2cm}
+(30+48p)\tau^2 +24\tau \big), \\
&& f_2(\tau)=\frac{1}{12}p(p+1)m \big( (-4p-5)\tau^6 +(4p+2)\tau^5 +(29+28p)\tau^4+(-16p-8)\tau^3 \nonumber \\
&&\hspace{2cm}-(18m+24p)\tau^2 \big).
\end{eqnarray}
\end{subequations} 

It follows then that
\begin{equation}
X^{-1}_{p+1,p}B_{p,p+1}= \frac{C_{p;p}}{X^0_{p+1,p}}
\begin{pmatrix}
\frac{\displaystyle (1+\tau)^{p-1}}{\displaystyle (1-\tau)^3}\big( J_{1}^s F_0 (1-\tau)^2 +(-1)^{p}J_3F_2
\big) \\
\frac{\displaystyle (1-\tau)^{p-1}}{\displaystyle (1+\tau)^3}\big( J_1 F_2 (1+\tau)^2 +
(-1)^{p}J_{3}^sF_0 \big)
\end{pmatrix}, 
\end{equation}
where the polynomials $J_1$ and $J_3$ arising from the fundamental matrix are
given by
\begin{subequations}
\begin{eqnarray}
&& J_1(\tau)=\frac{1}{2^{p+2}}P_{p-1}^{(p+2,-p)}(\tau), \label{J_1}\\
&& J_3(\tau)=\frac{1}{2^p} P_{p+1}^{(-p-2,p)}(\tau).   \label{J_3}
\end{eqnarray}
\end{subequations}
The polynomials $F_0$ and $F_2$ are given by
\begin{subequations}
\begin{eqnarray}
&& F_0=(1-\tau)^2 f_0 +(1+\tau)^2 f_2, \\
&& F_2=(1-\tau)^2 f_{2}^s +(1+\tau)^2 f_{0}^s,
\end{eqnarray}
\end{subequations}
 In order to determine whether the solutions of the transport equations at this
order contain logarithms, one has to decide, say, whether
\begin{equation}
H(\tau)=(1+\tau)^{p-1}\big( (1-\tau)^2 J^s_{1} F_0+(-1)^{p+1}J_3 F_2 \big) \label{H_0}
\end{equation}
and its derivatives have zeros at $\tau=1$. Direct evaluation of formula
(\ref{H_0}) using the identities listed in the appendix for the values
of the Jacobi polynomials at the terminal points $\tau=\pm1$ render
\begin{subequations}
\begin{eqnarray}
&& H(1)=2(1-p^2)m, \\
&& H'(1)=\frac{1}{3}(p+1)(5m-7pm), \\
&& H''(1)=0.
\end{eqnarray}
\end{subequations}

Using the division algorithm one can write
\begin{equation}
H(\tau)=(1-\tau)^2h_2(\tau)+(1-\tau)h_1(1)+H(1).
\end{equation}
where $h_1(\tau)$ and $h_2(\tau)$ are some polynomials. Thus,
$h_2(1)=0$, if $H''(1)=0$, whence
\begin{equation}
H(\tau)=(1-\tau)^3h_3(\tau)+(1-\tau)h_1(1)+H(1).
\end{equation}
Hence
\begin{equation}
X^{-1}_{p+1;p}B_{p+1;p}=\frac{C_{p;p}}{X^0_{p+1,p}} 
\begin{pmatrix}
h_3(\tau)+\frac{\displaystyle h_1(1)}{\displaystyle (1-\tau)^2}+\frac{\displaystyle H(1)}{\displaystyle (1-\tau)^3}\\
h^s_{3}(\tau)+\frac{\displaystyle h^s_{1}(1)}{\displaystyle (1+\tau)^2}+\frac{\displaystyle H^s(1)}{\displaystyle (1-\tau)^3}
\end{pmatrix},
\end{equation}
from where it follows that the solutions $a_{0,p+1;p,k}(\tau)$,
$a_{1,p+1;p,k}(\tau)$, $a_{2,p+1;p,k}(\tau)$ are polynomial as no
$(1\pm\tau)^{-1}$ terms have to be integrated. Note also, that the
multiplication of $\int X^{-1}_{p+1;p}B_{p+1;p}\mbox{d}\tau$ by $X_{p+1;p}$
cancels out all rational expressions. \hfill $\blacksquare$

\section{The order $p+2$}
It is now shown how the $\TT{2p}{k}{q-1+j}$ harmonics at order
$\mathcal{O}(\rho^p)$ in the initial data can give rise to solutions to
the transport equations with logarithmic divergences at the critical sets
$\mathcal{I}^\pm$. The main result is the following:

\begin{proposition} \label{proposition:p+2} 
There are initial data
  satisfying the regularity condition (\ref{regularity:condition})
  such that the functions $a_{0,p+2;p,k}(\tau)$, $a_{1,p+2;p,k}(\tau)$
  and $a_{2,p+2;p,k}(\tau)$ with $k=0,\ldots,2p$, solving equations
  (\ref{tp0}), (\ref{tp1}) and (\ref{tp2}) contain logarithmic
  divergences at $\tau=\pm 1$.
\end{proposition}

\textbf{Proof.} A look at the integral formula
(\ref{solution:integral}) reveals that the order $p+1$ solutions
$a_{0,p+1;p,k}(\tau)$, $a_{1,p+1;p,k}(\tau)$ and $a_{2,p+1;p,k}(\tau)$
are composed by two independent parts: one having to do with the
homogeneous part of the solution which can be eliminated by setting
$a_{0,p+1;p,k}(0)=a_{2,p+1;p,k}(0)$, $k=0,\dots,2p$ in the initial
data, and the other the nonhomogeneous solution proper which can be
``turned off'' by setting $a_{0,p;p,k}(0)=0$ ---assuming the
regularity condition (\ref{regularity:condition}) holds. For the
purpose of exhibiting the existence of logarithmic divergences in the
solutions to the order $p+2$ Maxwell transport equations it will
suffice to analyse what happens when only the \emph{homogeneous terms}
of the order $p+1$ solutions are considered ---exploiting the
linearity of the transport equations. Let $Y^h_{p+1;p,k}$ denote these
terms. From formula (\ref{solution:integral}) one has that
\begin{equation}
Y^h_{p+1}(\tau)=X_{p+1}(\tau)X^{-1}_{p+1}(0)\
\begin{pmatrix}
a_{0,p+1}(0) \\ a_{2,p+1}(0)
\end{pmatrix}, 
\end{equation} 
where again here and in the sequel the subindex string $_{;p,k}$ has
been suppressed. Consequently,
performing the replacement $p\rightarrow p+2$ and $q\rightarrow p$ in
the equations (\ref{new_tp0})--(\ref{new_tp2}) and setting
$a_{0,p}(\tau)=a_{1,p}(\tau)=a_{2,p}(\tau)=0$ ---which is the case if
$a_{0,p}(0)=0$--- one finds that the components $b_{0,p+2}$ and
$b_{2,p+2}$ of the column vector $B_{p+2}$ containing the
nonhomogeneous part of the system of ordinary differential equations
to be solved are given by
\begin{subequations}
\begin{eqnarray}
&& b_{0,p+2}=\frac{1}{(1+\tau)^3} \big(g_0 a_{0,p+1} + g_2 a_{2,p+1}\big), \\
&& b_{2,p+2}=\frac{1}{(1-\tau)^3} \big(g^s_0 a_{0,p+1} + g^s_2 a_{2,p+1}\big), 
\end{eqnarray}
\end{subequations} 
with the polynomials $g_0$ and $g_2$ given by
\begin{subequations}
\begin{eqnarray}
&& g_0(\tau)=\frac{m\tau}{12}(1+\tau)\bigg( (4p^2+9p)\tau^5
+(8p^2+10p-12)\tau^4
-(16p^2+33p-4)\tau^3 \nonumber \\
&& \hspace{3.5cm}-(56p^2+112p+8)\tau^2-(48p^2+126p+72)\tau-(24p+48)\bigg), \\
&& g_2(\tau)=-\frac{m\tau^2}{12}(1+\tau)\bigg((4p+9)\tau^4 -(4p+6)\tau^3
 \nonumber \\
&& \hspace{3.5cm} -(28p+57)\tau^2+(16p+24)\tau +(24p+42)   \bigg).
\end{eqnarray}
\end{subequations}
In the above expressions the order $p+1$ Maxwell transport equations have been
used to eliminate the derivatives $a'_{0,p+1}$, $a'_{1,p+1}$ and $a'_{2,p+1}$
that arise. Further, equation (\ref{auxiliary}) has also been used to dispose
of $a_{1,p+1}$.

For conciseness of the presentation, let
\begin{equation}
X_{p+1}^{-1}(0)Y_{p+1}(0)=
\begin{pmatrix}
z_0 \\
z_2
\end{pmatrix}.
\end{equation}
Hence, one has that
\begin{equation}
Y^h_{p+1}=
\begin{pmatrix}
a_{0,p+1}(\tau) \\
a_{2,p+1}(\tau)
\end{pmatrix}
=
\begin{pmatrix}
z_0(1-\tau)^{p+2} J_1(\tau) +(-1)^{p+1}z_2(1+\tau)^p J_3(\tau) \\
z_2(1+\tau)^{p+2} J^s_{1}(\tau) + (-1)^pz_0 (1-\tau)^p J^s_{3}(\tau)
\end{pmatrix}, \label{homogeneous:p+1}
\end{equation}
where $J_1$ and $J_3$ are the polynomials (\ref{J_1})--(\ref{J_3}) arising
from the order $p+1$ fundamental matrix.

As in the proof of proposition \ref{proposition:p+1}, the crucial
point in the argumentation is to analyse the decomposition in partial
fractions of the entries of
\begin{equation}
X^{-1}_{p+2,p} B_{p+2} =\frac{1}{X^0_{p+2,p}}
\begin{pmatrix}
\frac{\displaystyle 1}{\displaystyle (1-\tau)^{p+1}}(s_0 a_{0,p+1} + s_2 a_{2,p+1}) K^s_{1}
+\frac{\displaystyle (-1)^p}{\displaystyle (1-\tau)^{p+3}}(s^s_{0} a_{0,p+1} + s^s_{2} a_{2,p+1}) K_3 \\
\frac{\displaystyle 1}{\displaystyle (1+\tau)^{p+1}}(s^s_{0} a_{0,p+1} + s^s_{2} a_{2,p+1}) K_1 + \frac{\displaystyle (-1)^p}{\displaystyle (1+\tau)^{p+3}} (s_0 a_{0,p+1} +s_2 a_{2,p+1}) K^s_{3}
\end{pmatrix}, \label{big}
\end{equation}
where it has been used that $g_0=(1+\tau) s_0$ and $g_2=(1+\tau) s_2$,
and
\begin{subequations}
\begin{eqnarray}
&& K_1(\tau)=\frac{1}{2^{p+3}}P^{(p+3,-p-1)}_{p-1}(\tau),\\
&& K_3(\tau)=\frac{1}{2^{p+1}}(1+\tau)^{p+1}P^{(-p-3,p+1)}_{p+1}(\tau),
\end{eqnarray}
\end{subequations}
are the Jacobi polynomials arising in the fundamental matrix $X_{p+2,p}$. 
 
Contrary to what happens in the analysis of the order $p+1$, the functions
$a_{0,p+1}$ and $a_{2,p+1}$ are not homogeneous in $(1\pm
\tau)^m$ for some integer $m$. This fact complicates the analysis of
the decomposition in partial fractions of the expression (\ref{big}). However,
if one is interested in exhibiting the generic existence of logarithms in the
solutions, their appearance for a restricted type of initial data
suffices. With this idea in mind, set $z_2=0$ in equation
(\ref{homogeneous:p+1}) so that
\begin{equation}
a_{0,p+1}(\tau)= z_0(1-\tau)^{p+2}J_1(\tau), \quad a_{2,p+1}(\tau)= z_0 (-1)^p(1-\tau)^p J^s_{3}(\tau). 
\end{equation}
So, one finds that in particular
\begin{eqnarray}
&& \frac{1}{(1-\tau)^{p+1}}(s_0 a_{0,p+1} + s_2 a_{2,p+1}) K^s_{1} +\frac{(-1)^p}{(1-\tau)^{p+3}}(s^s_{0} a_{0,p+1} + s^s_{2} a_{2,p+1}) K_3 \nonumber \\
&& \hspace{2cm} = z_0 (1-\tau)K^s_{1} J_1 s_0  + \frac{G(\tau)}{(1-\tau)^3}, 
\end{eqnarray}
where
\begin{equation}
G=(-1)^p(1-\tau)^2 K^s_{1} J^s_{3} s_2  +(-1)^p (1-\tau)^2K_3 J_1 s^s_{0}  + K_3 J_3 s^s_{2}.
\end{equation}
As in the previous section, the problem reduces to calculating $G''$ and finding out whether it has zeros at $\tau=1$.

A direct calculation shows that
\begin{subequations}
\begin{eqnarray}
&& G(1)=p(p+2)^2 2^{-2p-1}m\binom{2p+1}{p+1}, \\
&& G'(1)=-\frac{m}{12}(p+2)(3p^3-4p^2-16p+2)\binom{2p+1}{p+1}, \\
&& G''(1)=\frac{m}{12}(p+2)(3p^4-4p^3-3p^2-36p-98)\binom{2p+1}{p+1}.
\end{eqnarray}
\end{subequations}
It can be seen that the only real positive root of the polynomial in
p, $3p^4-4p^3-3p^2-36p-98$, lies in the interval $(3,4)$ so that one
can conclude that $G''(1)\neq 0$, for $p\geq 1$. Using arguments
similar to those of the proof of proposition \ref{proposition:p+1} one
concludes that
\begin{equation}
G(\tau)= (1-\tau)^3 \tilde{h}_3(\tau)+(1-\tau)^2 \tilde{h}_2(1)+(1-\tau)\tilde{h}_1(1)+G(1),
\end{equation}
for some polynomials $\tilde{h}_1$, $\tilde{h}_2$ and $\tilde{h}_3$,
with $h_2(1)\neq 0$. Whence
\begin{equation}
\frac{G(\tau)}{(1-\tau)^3}=\tilde{h}_3(\tau)+\frac{\tilde{h}_2(1)}{1-\tau}+\frac{\tilde{h}_1(1)}{(1-\tau)^2}+\frac{G(1)}{(1-\tau)^3}.
\end{equation}
Thus, it follows that the integration in formula (\ref{solution:integral})
renders logarithmics terms $\ln(1\pm\tau)$. Note, again, that the
multiplication of the integral by $X_{p+2,p}$ cancels out all fractional terms
in $(1\pm\tau)$ so that the solution lies in the ring generated by polynomials
in $\tau$ and $\ln(1\pm\tau)$.
\hfill $\blacksquare$

\section{Conclusions}
The present work has been a first attempt to understand, in a general
way, the mechanisms logarithmic divergences arising at the critical
sets $\mathcal{I}^\pm$ in the solutions to the transport equations at
$\mathcal{I}$ implied by a Maxwell field propagating on a
Schwarzschild background. The ultimate goal behind this programme is
to develop ideas and methods which could be employed in the analysis
of the much more challenging case of the behaviour of the transport
equations implied by the Conformal Einstein equations. Preliminary
calculations show that by the methods employed in this article it
should be possible to at least obtain an analogue of propositions
\ref{proposition:p} and \ref{proposition:p+1} for this case. That is, starting from
initial data which are exactly Schwarzschildean up to order $p-1$, and
with a first deviation from Schwarzschild data occurring at order $p$,
it should be possible to fully integrate the order $p$ transport
equations ---note that in this case the system consists of 50
equations!--- to show that if a certain regularity condition in the data
holds ---the analogous of condition
(\ref{regularity:condition})--- then there are no logarithms in the
solutions of the transport equations at order $p$ and $p+1$. Quite
remarkably, some explicit calculations for concrete values of $p$ show
that there should not be any logarithms in the solution at order
$p+2$, and that the analogue of proposition \ref{proposition:p+2}
occurs in this case at order $p+3$. It is quite possible, however,
that a proof of these last two observations is beyond the
possibilities of the present methods, and that further ideas and
insights have to be developed.

In the light of the evidence gathered in this article one can not
avoid to feel that there is some powerful group theoretical structure
lurking behind which is the responsible of the
remarkable patterns that have been observed in both the case analysed
in this article and more generally in the conformal Einstein field
equations.

It would be of particular interest to reformulate the results obtained
here in a context or language more in tone with ideas and concepts of
Differential Galois theory, which in turn would render further
insights ---see e.g. \cite{Mag99} for a brief introduction to the
ideas and goals of Differential Galois theory. For example,
propositions \ref{proposition:p}, \ref{proposition:p+1} and
\ref{proposition:p+2} state that the Galois groups of the
order $p$ and $p+1$ transport equations are trivial, while that of the
order $p+2$ equations are not. Now, the question is the following: is
it possible to obtain these results by ---say-- looking directly at
the field equations and without having to solve almost explicitly the
equations?

\section*{Acknowledgements}
This research is funded by an EPSRC Advanced Research Fellowship. I
thank the Max Planck Institute for Gravitational Physics (Albert
Einstein Institute) for its hospitality during the course of a visit.
I thank CM Losert V-K for a careful reading of the manuscript.

\appendix

\section{Some spinors} \label{appendix:spinors}
Let $\{o_A, \iota_A\}$ denote normalised spinor dyads,
$\epsilon_{AB}o^A\iota^B=1$, where $\epsilon_{AB}$ is the standard alternating
spinor. In the text, the following spinors have been used:
\begin{subequations}
\begin{eqnarray}
&& \tau_{AA'}=o_A o_{A'} +\iota_A \iota_{A'}, \\
&& \epsilon^0_{AB}=o_A o_B, \quad \epsilon^1_{AB}=o_{(A}\iota_{B)}, \quad \epsilon^2_{AB}=\iota_A \iota_B, \\
&& x_{AB}=\frac{1}{\sqrt{2}}(o_A\iota_B+\iota_A o_B), \quad y_{AB}=-\frac{1}{\sqrt{2}}\iota_A\iota_B, \quad z_{AB}=\frac{1}{\sqrt{2}}o_A o_B, \\
&& \epsilon^0_{ABCD}=o_A o_B o_C o_D, \quad \epsilon^1_{ABCD}= o_{(A} o_B o_C \iota_{D)}, \quad \cdots, \quad \epsilon^4_{ABCD}=\iota_A \iota_B \iota_C \iota_D, \\
&& h_{ABCD}=-\epsilon_{A(C}\epsilon_{D)B}.
\end{eqnarray}
\end{subequations}

\section{The expansions of the Schwarzschild spacetime} \label{appendix:expansions}
The following expansions of the field quantities on the F-gauge for
the conformal Schwarzschild solution have been used in the discussion
of the article. The expansions have been calculated using the computer
algebra methods developed in \cite{Val04a}.
\begin{subequations}
\begin{eqnarray}
&& \hspace{-1.5cm}c^0=-\tau + \left(\frac{4}{3}m{\tau}^{3}-\frac{1}{3}m{\tau}^{5}
 \right)\rho + \frac{1}{2!}\left(\frac{1}{7}{m}^{2}{\tau}^{9}+{\frac {8}{7}}{m}^{2}{\tau}^{7}-3{m}^{2}\tau^5  -2{m}^{2}\tau^{3}  \right)\rho^2  +\O(\rho^3),  \\
&& \hspace{-1.5cm}c^1=\rho + \frac{1}{2!}\left(-4m{\tau}^{2}+\frac{2}{3}{\tau}^{4}
 \right)\rho^2 +\frac{1}{3!}\bigg( 12m^2\tau^2 +15m^2 \tau^4 -\frac{14}{3}{m}^{2}
{\tau}^{6}+\frac{3}{7}{m}^{2}{\tau}^{8}
 \bigg)\rho^3  +\mathcal{O}(\rho^4), \\
&&\hspace{-1.5cm} c^\pm= 1 + \left(m{\tau}^{2}-\frac{1}{6}m{\tau}^{4} \right)\rho +\frac{1}{2!}\bigg( \frac{1}{14}{m}^{2}{\tau}^{8}-{\frac {8}{9}}{m}^{2}{\tau}^{6}+ 3m^2 \tau^4-2m^2 {\tau}^{2} \bigg)\rho^2 +\mathcal{O}(\rho^3), \\
&& \hspace{-1.5cm}f= 1  + \frac{1}{3}m\tau^4 \rho +\frac{1}{2!}\left( \frac{1}{7}{m}^{2}{\tau}^{8}-\frac{4}{9}{m}^{2}{\tau}^{6}+ \frac{1}{3}{m}^{2}\tau^{4}-2{m}^{2} {\tau}^{2} \right)\rho^2 +\mathcal{O}(\rho^3), \\
&& \hspace{-1.5cm}\frac{\xi}{\sqrt{2}} = \left(\frac{1}{2}{\tau}^{2}m-\frac{1}{4}m{\tau}^{4}\right)\rho + \frac{1}{2!}\bigg( -\frac{1}{28}{m}^{2}{\tau}^{8}-\frac{2}{9}{\tau}^{6}{m}^{2}+\frac{4}{3}{m}^{2}\tau^{4} \bigg)\rho^2  +\mathcal{O}(\rho^3), \\
&& \hspace{-1.5cm}\chi_2 = \left(4m{\tau^3-12\tau m }
  \right)\rho + \frac{1}{2!}\left(  24{m}^2\tau -8{m}^{2}{\tau}^{3}+4{\tau}^{5}{m}^{2}-{\frac {4}{21}}{m}^{2}{\tau}^{7}
\right)\rho^2 +\mathcal{O}(\rho^3), \\
&& \hspace{-1.5cm}\chi_h= \frac{1}{2!}\left(-{\frac {20}{3}}{\tau}^{3}{m}^{2}+\frac{8}{3}{\tau}^{5}{m}^{2}-{\frac {20}
{63}}{m}^{2}{\tau}^{7}
  \right)\rho^2 +\mathcal{O}(\rho^3). 
\end{eqnarray}
\end{subequations}

For completeness the components of the curvature are also given:
\begin{subequations}
\begin{eqnarray}
&&\hspace{-2.5cm}\frac{\Theta_x}{\sqrt{2}}=4 m\tau
 \rho+ \frac{1}{2!}\left( \frac{4}{3}{\tau}^{5}{m}^{2}-\frac{8}{3}{\tau}^{3}{m}^{2} -12{m}^{2} \tau
 \right)\rho^2 +\mathcal{O}(\rho^3), \\
&&\hspace{-2.5cm} \Theta_2= 6m\left( 1-{\tau}^{2}
\right)\rho +\frac{1}{2!}\left( -12 {m}^{2}+12 {m
}^{2}{\tau}^{2}-10 {\tau}^{4}{m}^{2}+\frac{2}{3} {m}^{2}{\tau}^{6}
 \right)\rho^2 +\mathcal{O}(\rho^2), \\
&&\hspace{-2.5cm} \Theta_h= \frac{m^2}{2}\left(4 {\tau}^{2}-\frac{8}{3} {\tau}^{4}+\frac{4}{9}{\tau}^{6}
 \right) \rho^2 +\mathcal{O}(\rho^3), \\ 
&&\hspace{-2.5cm} \phi= -6m+ \left(-18 {m}^{2}{\tau}^{2}+3 {\tau}^{4}{m}^{2}
\right) \rho +\frac{1}{2!}\left(-{\frac {16}{7}}{\tau}^{8}{m}^{3}+28{m}^{3}{\tau}^{6}-90{m}^{3} \tau^{4}+36{m}^{3} {\tau}^{2} \right) \rho^2 +\mathcal{O}(\rho^3).
\end{eqnarray}
\end{subequations}
Finally, one has that
\begin{subequations}
\begin{eqnarray}
&& 2\Gamma_{0011}-4\Gamma_{1010}=2\xi+\frac{\sqrt{2}}{6}\chi_2-2\sqrt{2}\chi_h-\sqrt{2}f, \\
&& 2(\Gamma_{1100}+\Gamma_{0011})=-\frac{\sqrt{2}}{3}\chi_2-2\sqrt{2}\chi_h, \\
&& 2\Gamma_{1100}-4\Gamma_{0101}=-2\xi +\frac{\sqrt{2}}{6}-2\sqrt{2}\chi_h+\sqrt{2}f, \\
&& 2(\Gamma_{0011}-\Gamma_{1100})=3\xi -\frac{\sqrt{2}}{12}\chi_2-\frac{\sqrt{2}}{2}\chi_h.
\end{eqnarray}
\end{subequations}

\section{Some selected properties of the Jacobi polynomials} \label{appendix:jacobi}
The following properties of the Jacobi polynomials have been used in
the calculations described in this article. A good general reference to this
matter is the monograph \cite{Sze78}.

The Jacobi polynomials are defined by
\begin{equation}
P^{(\alpha,\beta)}_n(\tau)=\frac{\Gamma(\alpha+n+1)}{n! \Gamma(\alpha+\beta+n+1)}\sum_{m=0}^n \binom{n}{m}\frac{\Gamma(\alpha+\beta+n+m+1)}{\Gamma(\alpha+m+1)}\left(\frac{\tau-1}{2} \right)^m.
\end{equation}
In the above expression the combinatorial coefficients are understood in the generalised sense involving the Gamma function. That is,
\begin{equation}
\binom{m}{n}=\frac{\Gamma(m+1)}{\Gamma(n+1)\Gamma(m-n+1)}.
\end{equation}
Derivatives of the above can be rewritten as lower degree Jacobi
polynomials via
\begin{equation}
\frac{\mbox{d}^k}{\mbox{d}\tau^k}P^{(\alpha,\beta)}_n(\tau)=\frac{\Gamma(\alpha +\beta+n+1+k)}{2^k\Gamma(\alpha+\beta+n+1)} P^{(\alpha+k,\beta+k)}_{n-k}(\tau).
\end{equation}
The polynomials satisfy the symmetry relation
\begin{equation}
P^{(\alpha,\beta)}_n(-\tau)=(-1)^nP_n^{(\beta,\alpha)}(\tau).
\end{equation}
The value of the polynomials at $\tau=\pm 1$ is given by
\begin{equation}
P^{(\alpha,\beta)}_n(1)=\binom{n+\alpha}{n}, \quad P^{(\alpha,\beta)}_n(-1)=(-1)^n\binom{n+\beta}{n}.
\end{equation}


\end{document}